\input epsf
\documentstyle[12pt]{article}
\begin{document}
\hoffset=-1.5cm
\hsize=165truemm
\tolerance=5000
\def\phrac#1/#2{\leavevmode\kern.1em\raise.5ex\hbox{\the\scriptfont0 #1}
\kern-.1em/\kern-.15em\lower.25ex\hbox{\the\scriptfont0 #2}}
\def \lp{\scriptscriptstyle +}
\def \lm{\scriptscriptstyle -}
\def\sss{\rm_S}
\def\PARISC{\small PA-RISC}
%
\baselineskip=18 pt plus 2pt minus 1pt
\itemsep 3pt
\parskip  5pt plus 1pt
\centerline{\bf Using Multiple RISC CPUs in Parallel
to Study Charm Quarks}
\vskip 15pt
\centerline{by}
\centerline{C. Stoughton}
\centerline{Fermilab, Batavia, Illinois 60510 \ USA}
\vskip 10pt
\centerline{and}
\vskip 5pt
\centerline{D. J. Summers}
\centerline{Department of Physics and Astronomy}
\centerline{University of Mississippi, Oxford, MS 38677 \ USA}
\vskip 25pt
\centerline{\bf ABSTRACT}
\vskip 5pt
  We have integrated a system of 16 RISC CPUs to help reconstruct and
analyze a 1.3 Terabyte data set of 400 million high energy
physics interactions. These new CPUs provided an affordable means of processing
a very large data set.
The data was generated using a
hadron beam and
a fixed target at Fermilab Experiment 769. Signals were
recorded on tape from particles created in or decaying near the target
and passing though a magnetic spectrometer. Because all the
interactions were independent, each CPU could completely reconstruct
any interaction without reference to other CPUs. Problems of this sort are
ideal for multiple processors. In the offline reconstuction system, we
used Exabyte 8mm video tape drives with an I/O capacity of 7 Terabytes
per year and a storage capacity of 2.3
Gigabytes per tape. This reduced tape mounts to one or two per day
rather than one or two per hour as would be the case with 9-track tapes.
The ETHERNET$^{TM}$ network used
to link the CPUs and has an I/O capacity of 15 Terabytes per year. The
RISC CPUs came in the form of commercially supported workstations with little
memory and no graphics to minimize cost.  Each 25 MHz MIPS R3000 RISC
CPU processed data 20 times faster than 16MHz Motorola 68020 CPUs that
were also used. About 8000 hours of processing was needed to
reconstruct
the data set.  A sample of thousands of fully reconstructed particles
containing a charm quark has been produced.
\vskip 20pt
\leftline{\bf I. INTRODUCTION}
\vskip 3pt
 
    The computing needs of many experiments in high energy particle physics can
be met using multiple CPUs working in parallel.  Typical experiments record
$10^6$ to $10^{11}$ independent events.  The results of the computation
performed on one event do not affect other events.  Therefore, it is straight
forward to adapt these problems to a parallel processing environment [1].
 
    Experiment 769 [2] at the Fermi National Accelerator Laboratory (Fermilab)
studies the production of particles containing the charm quark.   During this
experiment, 400 million interactions were recorded at the Tagged Particle
Spectrometer [3] on 9000 nine track tapes. The system of UNIX workstations that
we describe here performs two compute intensive tasks on this data set.  First,
the event reconstruction algorithm, which requires three-fourths of a CPU
second per
event, reconstructs particle trajectories, momenta, and type.  Second, a
filtering algorithm inspects each event and retains candidates that could
contain charm particles. We could have performed both tasks at once but
chose instead to write all of the reconstructed events on serial media during
the first pass of the data.  Then, we read the reconstructed events and filtered
during the second pass.
 
   High energy particle physics is often similar to gold mining.  A miner
sifts through an enormous amount of rock to find specks of gold.  A physicist
often has to examine an enormous number of particle interactions to find rare
events. In E769, a beam of 250 GeV/c pions,
kaons, and protons interacted with 26 metal foils.  A 9 month run produced a
yield of over 2 billion interactions,
of which 400 million were selected and recorded.
Out of these interactions, we have been
able to reconstruct thousands of particles containing the charm quark [4].
Typical decays include the two modes shown in Figure 1,
\vskip 3pt
\centerline{$D^0(c\overline{u}) \rightarrow
K^-(s\overline{u})\; \; \pi^+(u\overline{d})$ and}
\centerline{$D^+(c\overline{d}) \rightarrow
K^-(s\overline{u})\; \; \pi^+(u\overline{d})
\; \; \pi^+(u\overline{d})$.}
 
   Charm particles are relatively massive and long lived.  We calculated the
mass from the measured four-momentum of decay particles.  The long charm
lifetime leads to a decay length of a few millimeters.
The decay length in each event is the difference
between two positions:  the location of the
interaction of the beam particle with the target foil and the location of the
decay of the charm particle.  If the mass and decay length
were calculated as each
event happened, it would be easy to select only the few thousand
containing charm quarks.  However,
it is very expensive to calculate these quantities in real time.  We
found it easier to selectively record 400 million events out
of 2 billion
interactions with quantities constructed directly from the analog signals
of the detector.  This selection was based on the amount of energy that
is produced transverse to the beam direction, since events containing charm
quarks will have more transverse energy than the more frequent events with
the lighter up, down, and strange quarks.
This selection yielded 1.3 terabytes of event data, which were written on
9,000 nine track tapes with a high bandwidth data acquisition system [5].
We later copied these data to the 8mm format.  The rapidly improving
price/performance of some types of offline computing allowed us to play back
these tapes and completely reconstruct the events.
 
\vskip 15pt
\leftline{\bf II. SYSTEM REQUIREMENTS}
\vskip 3pt
 
   We wanted to speed the extraction of charm events from this data. To do
this a system of Reduced Instruction Set Computer
(RISC) processors was devised. This supplemented a large number of Fermilab
Advanced Computer Program Motorola 68020
microprocessors (ACP-I) [6] already shared by many Fermilab experiments.
The idea was to build a dedicated system for one project which would
avoid the delays and expense of a general purpose computer
which could support scores of
software packages and users.
 
  An in-depth market survey was performed and then vendors were asked to bid
on a
fully competitive basis.  This led to the purchase of one Silicon Graphics (SGI)
[7] 4D/240S compute server by the Fermilab Physics Section which was used by
E769 and other experiments.  
Three additional SGI 4D/240S
compute servers dedicated to E769 were subsequently purchased.
The network architecture of this system
of compute servers is shown
in Figure 2.
 
  Each SGI 4D/240S utilizes four 25 MHz processors developed by
MIPS Computer Systems, the MIPS R3000 RISC CPU and
MIPS R3010 Floating Point Unit.
The four processors in one of the 4D/240S servers share 16MB
of memory.  The other three servers each have 8MB of shared local memory.
A dedicated bus
is used to connect processors and memory within the compute server.
We tested the performance of the shared memory by measuring the performance
of a task when it is running alone, and then with three other tasks on the
compute server.
A performance degradation of a
few per cent was noticed in the first job run on the 4D/240S when three more
jobs ran concurrently.  Characteristics of the processor and memory systems are
in Table 1.    Table 2 compares its performance
to Fermilab's ACP-I
Motorola 68020 processors.  We had nine key goals.
The SGI systems solved our nine goals as follows:
 
\newcounter{bean}
\begin{list}
{\arabic{bean}.}{\usecounter{bean} \setlength{\rightmargin}{-1.5cm}}
\item  {\bf GOAL.} We needed enough CPU power to complete
   the event reconstruction in a year.
   Additional delays would dilute the scientific interest in the results
   of this data set.  We needed the processing power of an additional 300
   Motorola 68020 CPUs.  Our goal was get this processing power and to
   simultaneously minimize the cost of reconstructing an event.
 
   {\bf SOLUTION.} The SGI system
   provided the raw CPU power that we needed to find charm quarks  in a
   timely fashion at a price about 50 times lower than a traditional mainframe,
   and three times lower than the cost of additional ACP-I systems.
   Anything that needlessly added cost (e.g. graphics displays or
   extra memory) was rejected.

\item {\bf GOAL.} Commercial availability:  We wanted to minimize designing,
   building and maintaining the computer
   hardware and operating system.
 
   {\bf SOLUTION.} The SGI system
   was commercially supported and incurred little in the way of
   engineering and self-maintenance costs.  A three year software and hardware
   maintenance contract was bundled with each compute server.
 
\item {\bf GOAL.}
   We needed robust, optimizing FORTRAN and C compilers.
   It would have been difficult
   to translate the 58,000 lines of the E769 FORTRAN reconstruction program into
   assembly code by hand.  The assembly code would have been nearly
   impossible to maintain even if it could be produced.  We also wished to
   avoid debugging a FORTRAN compiler.
 
   {\bf SOLUTION.} The MIPS corporation
   took the unusual approach of developing their high level
   language compilers and hardware in parallel. RISC
   compilers in general must be able to overlap instructions in order
   to take advantage of the hardware.
   For example, during a MIPS floating point divide 11 other instructions
   can also be executed [8].  MIPS
   brought out excellent FORTRAN
   and C compilers at the same time they brought out their RISC chips.
   SGI provided and supported versions of these compilers.
   FORTRAN licenses were purchased for the two compute servers that were
   used for software development.
 
\item  {\bf GOAL.} Data Throughput:
   Moving 1.3 terabytes of input data and an equal amount
   of output data to and from processors over a period of a year
   requires an average I/O rate of 85 kilobytes per second.  The data had to
   be read from the input media, distributed to the processors, collected,
   and written to the output media.
 
   {\bf SOLUTION.} The data
   throughput was provided by Exabyte 8mm tape drives [9]
   for the input and output data streams. ETHERNET$^{TM}$
   provided a network data path between servers.
   Small Computer System Interfaces (SCSI) were used to move data to and
   from the tape drives.
   The I/O rate of an Exabyte 8mm drive is 210 kByte/sec on the SGI 4D/240S,
   about
   a factor of five faster than necessary for an E769 reconstruction input tape.
   For a comparison of various media characteristics see Table 3.
   A single ETHERNET$^{TM}$ can move 15 terabytes per
   year given a rate of 0.5 MB/s.  This exceeded our minimal I/O requirement
   by a factor of six.  Typically a CPU processed a
   4kb data event for $\phrac3/4$ of a second.  During this time no I/O
   was required.
 
\item {\bf GOAL.} Convenience:  For operational simplicity, we wanted to
   limit tape mounting to a few short periods per day.  This
   avoids having to run shifts 24 hours per day to load tapes.
 
   {\bf SOLUTION.} If we had used nine track tapes directly, the system of
   compute servers would
   have required over 30 tape mounts per day to process 6,000 input and
   output tapes over a year.
   To avoid this bottleneck, we used the Fermilab tape copy facility to
   transfer data to the 8mm media.  Each 8mm tape can hold 2.3 Gigabytes of
   data, or up to 13 nine track tapes.
   This decoupled the large number of nine track tape mounts from the continuous
   data flow required by the compute servers and reduced the media cost for the
   reconstructed data by a factor of 25.
   We worked with SGI to adapt the Exabyte 8mm tape drives to the 4D/240S.
   Our data format has variable length blocks up to 65 kilobytes.  Before
   purchasing the system we copied
   an E769 nine track data tape to an 8mm cartridge.
   By reading this tape on
   an SGI system we verified that the software drivers included in the
   operating system
   handle large variable length blocks.  We purchased the first tape drive from
   SGI to ensure system integrity. The remaining drives were supplied by
   third party vendors.
 
\item {\bf GOAL.} Software to distribute the events to CPUs and collect results:
   The compute tasks can be performed
   in parallel by having a complete copy
   of the reconstruction or filtering algorithm on each of many parallel CPUs.
   The data flow, process scheduling, and bookkeeping tasks require careful
   software design.
 
   {\bf SOLUTION.} The Fermilab Advanced Computer Program's Cooperative
   Process Software (CPS) [10] was used to
   distribute events and collect results.  This was the first use
   of CPS in a physics experiment and it was well supported.
 
\item {\bf GOAL.}
   Software to communicate with VAX/VMS computers:  The Fermilab Physics
   Section's Local Area VAX Cluster (LAVC)
   is an interactive system of VAXstations which supports
   the general computing needs of physicists at Fermilab. The LAVC is shown in
   Figure 2.
   The FORTRAN code for our
   reconstruction and filtering algorithms was developed and maintained on
   this cluster.
 
   {\bf SOLUTION.} The SGI computers support TCP/IP communications.
   Multinet [11]
   software was in use at Fermilab, and this was installed on the VAX/VMS
   systems that we used, to provide them with TCP/IP communications.
 
\item {\bf GOAL.} Sufficient memory for efficient CPU use:
   1.3 MB of memory was needed per CPU during execution
   to contain the E769 reconstruction
   program and required data without paging to disk.
 
   {\bf SOLUTION.} A total
   of 8 MB of memory was put on three of the SGI computers and 16 MB
   on the fourth.  The 8MB computers thus had 2 MB per CPU which allowed
   the 1.3 MB E769 reconstruction program to run without paging, even after the
   operating system's memory usage.
   Excessive memory can add substantially to the
   cost of a system.  To minimize our memory requirements the
   reconstruction and filtering algorithms were run as
   separate passes through the data.
 
\item {\bf GOAL.} We needed about
   5 gigabytes of disk to store user programs and
   intermediate data.
 
   {\bf SOLUTION.} We wished to try to reserve the
   bandwidth of the SCSI buses for Exabyte
   8mm tape drives.
   To do this we had SGI put one four-channel Enhanced Small Disk
   Interface (ESDI) in three of the compute servers and two in the fourth.
   As shown in Figure 2 all system disks are EDSI as well as seven user
   disks on the first  compute server.
   Five SCSI disks are also used.  All disks are $5\, \phrac1/4''$.
   Many of the disks
   included a 5 year warranty.  This greatly lowers
   the life cycle cost of a disk.  Network File System (NFS$^{TM}$) software is
   used to share disks between the different compute servers.
\end{list}
\vskip 15pt
\leftline{\bf III. SOFTWARE}
\vskip 3pt
 
    The compute servers ran the IRIX operating system, which is the
Silicon Graphics implementation of UNIX, based on System V.3 with BSD
4.3 enhancements.  IRIX is fully symmetric, allowing each process to
run on any of the processors within one of the compute servers.
The IRIX operating system managed CPU scheduling automatically.
 
    A good FORTRAN-77 compiler was an essential software tool for our
application, since the algorithms contain 58,000 lines of
FORTRAN.  The MIPS FORTRAN compiler
is reliable, and we had little trouble porting our
code from the VAX/VMS environment.  The three optimization levels are
--O0 (no optimization), --O1 (default, basic optimizations), and --O2
(global optimization).  In Table 4 we compare the performance of one
R3000 processor at these three optimization levels.  The source line
debugger (dbx) and profiler (pixie) utilities were useful in porting
and optimizing this code.
 
    We kept the sixteen processors in this system busy by using the
CPS package mentioned above to
distribute events among the processors.  The structure of one job is
shown in Figure 3.  Each of the bubbles represents one process on a
compute server.  There were three different kinds of tasks:  input, output, and
compute.  The input and output tasks were each carried out by a
single process.
The compute task was either the event reconstruction algorithm or the event
filtering algorithm.  This compute-intensive part of
the application was carried out by sixteen processes distributed
among the four compute servers.  One additional process, the job
manager, synchronized the work of these processes.
 
    The input process read the data stream from tape and sent blocks
of events to each compute process.  It then waited until one of
the compute processes became available.  The compute
process received the event block, performed the event reconstruction
or event filtering
algorithm to generate an output event block, sent this block to
the output process, and then signaled the input process that it was
available.  The output process polled the compute processes,
received data from each process as it become available, and wrote it
to tape or disk.
 
    To keep the system busy full time while we were running the reconstruction
algorithm, we used two pairs of tape drives.
One pair had the input and output tapes for the active job,
and the second pair of drives contained the tapes for the next
job.  The physicist on shift who determined which data to process and
monitored the progress of the event reconstruction needed to attend to the
system only two or three times daily.  This avoided the costs
associated with 24-hour operator coverage, and kept the system
utilization above 95\%, comparable to batch mainframe utilization.
System failures were infrequent and caused no
appreciable loss of time.

After the event reconstruction was completed we used the same system
to reduce the data sample from 400 million reconstructed events to the few
thousand events containing charm quarks.  Filtering algorithms select events
with charm quarks and reject events that do not contain charm quarks.  We
performed this filtering in two passes.  In the first pass
we reduced the data set with a general filter algorithm.
It selected events containing a pair of particle
tracks intersecting at a point downstream of the primary interaction point.
This first stage reduced the number of events by a factor of 15, leaving
just under 30 million events for the second filtering pass.

The most significant difference between the event reconstruction
task and the filtering task was the number of compute cycles required
for each event.  The characteristics of the two jobs are summarized in
Table 5.
For the filtering task, we were not able to keep all sixteen of the
processors busy with a single input data stream, since the I/O rate
is limited to 210 kByte/sec for a single Exabyte 8mm tape drive.
We split the filtering task into two concurrent data streams,  using
each stream to feed eight processors.  The  ETHERNET$^{TM}$ connection
between servers was able to handle the aggregate bandwidth of 292
kByte/sec during filtering.

A second, and related, difference was with tape mounts.  Since each 8mm
cartridge was filtered in three hours (instead of the ten hours required
for event reconstruction) we used additional tape drives to hold the tapes
waiting for execution.  With all the tape drives loaded, the system ran the
filtering task unattended for over 12 hours.  A final difference was with the
output data stream.  Since the amount of data was reduced by a factor of 15 in
the filtering task, we were able to write the output to separate disk files.  
Once a day, these files were copied to an output tape to free up disk space.

After the first stage of filtering the reduced data set fit on
23 8mm cartridges.  The next stage of data filtering
used criteria specific to the different charm particles and decay modes
that we are studying.  We ran different filtering algorithms on the reduced
data set to extract the final event samples.  For these final filtering
stages we have been using an automatic tape loader [12]
which has two Exabyte 8mm tape drives
and slots for 54 8mm cartridges in a carousel.
These stages of the data filtering algorithms also ran in the CPS environment.
The automatic tape mounts allowed us to scan and filter
the complete data set without intervention.
Different teams of physicists have developed, tested, and run new
filtering algorithms on the complete filtered data set in a matter of days.

%
%
 
\vskip 15pt
\leftline{\bf IV.  CONCLUSIONS}
\vskip 3pt
 
  One year after the arrival of the 16 RISC CPUs in July 1989, the
reconstruction of the 400 million E769 events was completed, and the physics
analysis [4] of thousands of particles with charm quarks was well under way.
This was the first time UNIX/RISC computers were used for such a large data set
in High Energy Physics. The affordable compute power in workstations could be
exploited because our processing can be broken down into tasks which process
independent events. The I/O bandwidth of ETHERNET$^{TM}$ was sufficient to
distribute events. The operational expense of running shifts 24 hours a day to
load tapes was avoided by using Exabytes.
 
  The 58,000 line program used to reconstruct E769 data is now serving as a
benchmark to track the rapidly improving cost performance of workstations (see
Table 6). Given this rapid progress, it is often prudent to buy a system {\it
just in time} and bring it on-line quickly.
 
\vskip 15pt
\leftline{\bf V. ACKNOWLEDGEMENTS}
\vskip 3pt
 
  We thank D.\ Green, L.\ Lederman, C.\ Kastner, S.\ Bracker, C.\ Johnstone,
M.\ Isely, J.\ Pfister, R.\ Sidwell, L.\ Lueking, M.\ Streetman, G.\ Luste,
B.\ Yeltsin, and P.\ Karchin
for their contributions to this effort.
The majority of this work was performed at the
Fermi National Accelerator Laboratory,
which is operated by Universities Research Association, Inc., under
contract DE-AC02-76CHO3000 with the U.S. Department of Energy.
This work was also supported by the
U.\ S.\ Department of Energy (DE-FG05-91ER40622) and
the National Research Council of Canada.
\def\issue(#1,#2,#3){\space$\underline{#1}$\space(#2)\space#3}
\def\PRL(#1,#2,#3){ Phys. Rev. Lett.\issue(#1,#2,#3)}
\def\PL(#1,#2,#3){ Physics Letters B\issue(#1,#2,#3)}
\def\PLB(#1,#2,#3){ Physics Letters\issue(#1,#2,#3)}
\def\PR(#1,#2,#3){ Phys. Rev.\issue(#1,#2,#3)}
\def\NC(#1,#2,#3){ Lettere Al Nuovo Cimento\issue(#1,#2,#3)}
\def\NIM(#1,#2,#3){ Nucl. Inst. Meth.\issue(#1,#2,#3)}
\def\NP(#1,#2,#3){ Nuclear Physics\issue(#1,#2,#3)}
\def\ZP(#1,#2,#3){ Zeitschrift f\"ur Physik C\issue(#1,#2,#3)}
\def\IEEE(#1,#2,#3){IEEE Trans. Nucl. Sci.\issue(#1,#2,#3)}
\eject
\leftline{\bf REFERENCES}
\smallskip
\begin{list}
{[\arabic{bean}]}{\usecounter{bean} \setlength{\rightmargin}{-1.5cm}}
\item [$^{TM}$] ETHERNET is a trademark of the Xerox Corporation. \quad
      NFS is a trademark of SUN Microsystems.
 
\item  Pioneering work in the use of multiple processors has been done by
       the Fermilab ACP group [6] and also at SLAC
       (see Paul F. Kunz et al., Experience using the 168/E Microprocessor
       for Off-line Data Analysis, \IEEE(27,1980,582)).
 
\item  E769 Collaborators. \quad   CBPF-Rio de Janeiro,
       Fermilab, Mississippi, Northeastern, Toronto, Tufts, Wisconsin, and Yale.
 
\item   B.\ H.\ Denby et al.\ (E516), Inelastic and Elastic
        Photoproduction of J/$\psi$(3097), \PRL(52,1984,795);
 
        J.\ R.\  Raab et al.\ (E691), Measurement of the
        $D^+$, $D^0$, and $D_{\sss}^{\lp}$
        Lifetimes, \PR(D37,1988,2391);
 
        J.\ C.\ Anjos et al.\ (E691),
        Measurement of D$_{\sss}^+$ Decays and Cabibbo-Suppressed D$^+$ Decays,
        \PRL(60,1988,897);
 
        J.\ C.\ Anjos et al.\ (E691),
        A Study of $D^0$--$\overline{D^0}$ Mixing from Fermilab E-691,
        \PRL(60,1988,1239);
 
        J.\ C.\ Anjos et al.\ (E691),
        Measurement of the $\Lambda_c^+$ Lifetime,
        \PRL(60,1988,1379);
 
        J.\ C.\ Anjos et al.\ (E691),
        Measurement of the Form-Factors in the Decay,
        $D^+ \rightarrow \overline{K}^{*0} e^+ \nu_e$,
        \PRL(65,1990,2630).
 
\item  D. Errede et al., Hadroproduction of Charm at Fermilab E769,
       $25^{th}$ International Conf.\ on High Energy
       Physics, Singapore, 2-8 Aug.\ 1990.
 
\item  C. Gay and S. Bracker, \IEEE(34,1987,870); \newline
       R. Vignani et al., \IEEE(34,1987,756);     \newline
       S. Hansen et al., \IEEE(34,1987,1003);     \newline
       Mark Bernett et al., \IEEE(34,1987,1047);  \newline
       R. Hance et al., \IEEE(34,1987,878).
 
\item  I. Gaines et al., The ACP Multiprocessor System at Fermilab,
       Proc.\ of the International Conf.\ on Computing in High Energy
       Physics, Asilomar, USA (2--6 Feb.\ 1987) 323; \quad FERMILAB-Conf-87/21.
 
       J. Biel et al., Software for the ACP Multiprocessor System,
       Proc.\ of the International Conf.\ on Computing in High Energy
       Physics, Asilomar, USA (2--6 Feb.\ 1987) 331; \quad FERMILAB-Conf-87/22.
 
\item
     Silicon Graphics (SGI), 2011 N Shoreline, Mountain View, CA 94039 \newline
     Digital Equipment (DEC), 146 Main St., Maynard, MA 01754    \newline
     Sony Microsystems, 645 River Oaks Parkway, San Jose, CA 95134 \newline
     MIPS Computer Systems, 950 DeGuigne Dr., Sunnyvale, CA 94086  \newline
     Hewlett-Packard/Apollo, 3000 Hanover St., Palo Alto, CA 94304 \newline
     IBM, Research Park, Suite 106, Starkville, MS 39759         \newline
     SUN Microsystems, 2550 Garcia Ave., Mountain View, CA 94043 \newline
     Cray Research, 655 Lone Oak Dr., Eagan, MN 55121
 
\item  Gerry Kane, MIPS RISC Architecture, Prentice Hall (1989) 7--12.
 
\item  Exabyte Corp., 1745 38th Street, Boulder, CO 80301.
 
\item  Joe Biel, Mike Isely, Mark Fishler, Chip Kaliher and Matt Fausey,
       ACP Cooperative Processes User's Manual, Fermilab GA0006, 19 Nov.\
       1990.
 
\item TGV Inc., 603 Mission St., Santa Cruz, CA 95060
 
\item IGM Data Autoloader Division, 4041 Home Rd., Bellingham, WA 98226
 
\item  System Performance Evaluation Cooperative (SPEC) c/o Waterside
       Associates, 39510 Paseo Padre Parkway, Suite 350, Fremont, CA 94538.
\end{list}
\eject
\leftline{\bf TABLES}
$$ \vbox {\halign{#\hfil& #\hfil &\qquad # \hfil \cr
Processor: &  Central Processor &  R3000                             \cr
           &  FP Processor      &  R3010                             \cr
           &  FP Data Format    &  IEEE 754, 32-- and 64--bit formats  \cr
           &  Registers         &  32 CPU, 32 FP (single precision)  \cr
           &  Word Length       &  32 bits                           \cr
           &   Clock Speed      &  25 MHz                            \cr
           &                    &                                    \cr
Cache Memory:& Cache Type        &  Write-back                        \cr
             & Cache Size        &  64 KB instruction                 \cr
             &                   &  64 KB Data (1st level)            \cr
             &                   &  256 KB Data (2nd level)           \cr
             & Write Buffer      &  4 words deep                      \cr
             & Read Buffer       &  16 words deep                     \cr
             & Processor Bus BW  &  64 MB/sec sustained               \cr
             &                   &                                    \cr
CPU Memory: & Size              &  8MB (16 MB on one server)         \cr
            & Maximum allowed   &  128 MB                            \cr
            & Bandwidth         &  64 MB/sec sustained               \cr
            &                   &                                    \cr
Virtual Memory: &               &  2 GB per process                  \cr
                &               &                                    \cr
MPLink Bus: & Width             &  32 bit address; 64 bit data       \cr
            & Bandwidth         &  64 MB/sec sustained               \cr
}
\vskip 10pt
\noindent
{\bf Table 1.  Characteristics of Silicon Graphics 4D/240S processors.} \quad
Each SGI 4D/240S contains four R3000 central processing units
and four R3010 floating point units with local caches and shared main memory.
}$$
\eject
\baselineskip=14 pt plus 1pt minus 1pt
$$ \vbox {\halign{#\hfil&\quad #\hfil&\quad #\hfil&\quad \hfil#\hfil
&\quad #\hfil
&\quad #\hfil&\quad \hfil# \cr
        & CPU        & CPU/FPU    & Clock & Instruction & Data  & E769     \cr
        & Maker      &            & MHz   & Cache       & Cache & sec/event \cr
        &            &            &       &             &       &           \cr
ACP-I       & Motorola  & 68020/68881 & 16    & 256 bytes &   0     & 18.01 \cr
ACP-I       & Motorola  & 68020/68882 & 16    & 256 bytes &   0     & 15.32 \cr
SGI 4D/240S & MIPS      & R3000/R3010 & 25    & 64kb      & 320kb   &  0.76 \cr
}}$$
\smallskip
\noindent
{\bf Table 2.  A comparison of SGI 4D/240S and Fermilab ACP I
computers.} \quad
The time
shown is that required to run the E769 particle reconstruction benchmark
on one CPU.  The Motorola 68882 is a faster version of the Motorola 68881
floating point chip.  Concurrent execution of some floating point
instructions are allowed in the 68882 and not allowed in the 68881.
\bigskip
\bigskip
\bigskip
$$ \vbox {\halign{#\hfil&\quad \hfil#&\quad #\hfil&\quad #\hfil&\quad \hfil#
&\quad \hfil# \cr
Tape Type & Length  & Capacity & \$/tape      & \$/Terabyte     & Tapes/ \cr
          &         &          &       &                        & Terabyte \cr
8mm Video &  106m   & 2.3  GB  & \$4.25       & \$ 1848      & 435           \cr
4mm DAT   &   60m   & 1.2  GB  & \$7.79       & \$ 6492      & 833           \cr
IBM 3480  &  165m   & 0.22 GB  & \$4.60       & \$20909     & 4545          \cr
9-track  &  732m   & 0.16 GB  & \$9.31       & \$58188     & 6250          \cr
}}$$
\smallskip
\noindent
{\bf Table 3.  A Comparison of Storage Media.} \quad
The 8mm, 9-track, and 3480 tape prices are from the Fermilab stockroom
catalog. The 4mm DAT price is from the New York Times, 20 Jan.\ 1991,
page 31.
\bigskip
\bigskip
\bigskip
$$ \vbox {\halign{#\hfil&\qquad #\hfil \cr
         Optimization Level:     &       Time: \cr
                                 &             \cr
         --O0 (no optimization)  &       1.45 seconds/E769 event \cr
         --O1 (default optimization) &   1.09 seconds/E769 event \cr
         --O2 (global optimization)  &   0.76 seconds/E769 event \cr
}}$$
\smallskip
\noindent
{\bf Table 4.  MIPS R3000 at different optimization levels.}
The time
shown is that required to run the E769 particle reconstruction benchmark
on a single CPU of an SGI 4D/240 computer.
\bigskip
\eject

\smallskip
\begin {tabular}{lll}
Processing Task                 & Reconstruction        & Filtering     \\
CPU Time/Event                  & 0.76 seconds          & 0.07 seconds  \\
\# Input Events                 & 400 $\times 10^6$     & 400 $\times 10^6 $ \\
Average Input Event Size        & 3.2 kByte             & 1.2 kByte \\
\# Processors per Data Stream   & 16                    & 8 \\
Input Bandwidth                 & 67 kByte/sec          & 137 kByte/sec \\
\# Output Events                & 400  $\times 10^6$    &  27 $\times 10^6$ \\
Average Output Event Size       & 1.2 kByte             & 1.2 kByte \\
Output Bandwidth                & 25 kByte/sec          & 9 kByte/sec \\
\# Concurrent Data Streams      & 1                     & 2 \\
Aggregate Bandwidth             & 92 kByte/sec          & 292 kByte/sec
\end {tabular}
\smallskip

{\bf Table 5.  Comparison of event reconstruction and filtering tasks.}
The throughput requirements are more demanding for the filtering
task, so we used two independent data streams to keep the entire system
efficiently utilized.

$$\vbox
{\halign{#\hfil&\ #\hfil&\hfil#&\ \hfil#\hfil&\ \hfil#
&\ \hfil#
&\ #\hfil&\ #\hfil&\ #\hfil  \cr
               &       &     & Cache & Write  &Main  &       &      &E769    \cr
               &       & CPU & Memory& Through&Memory& SPEC- & E769 &Normal- \cr
 COMPUTER [7] & CPU  &Clock& I+D+2nd & Buffer &Speed.& marks & sec/ &ized to \cr
               & Type  & MHz & kByte & Depth  &MB/sec& [13]  & event&25MHz   \cr
               &       &     &       &        &      &       &      &      \cr
 SGI Crimson S & R4000 &  50 & 8+8+1024 & N/A & 400  & 70.   &      &      \cr
 SGI 4D/35S    & R3000 &  36 & 64+64 &  11    &  97  & 31.   & 0.53 & 0.76 \cr
 SGI INDIGO    & R3000 &  33 & 32+32 &  11    &  91  & 26.   & 0.60 & 0.79 \cr
 SGI 4D/25S    & R3000 &  20 & 64+32 &   1    &   7  & 14.   & 1.07 & 0.86 \cr
 SGI 4D/20     & R3000 &  12 & 16+ 8 &   1    &   7  &  8.   & 1.94 & 0.93 \cr
 DEC 3100      & R2000 &  16 & 64+64 &   4    &  13  & 11.7  & 1.10 & 0.73 \cr
 DEC 5000-200  & R3000 &  25 & 64+64 &   6    & 100  & 23.5  & 0.66 & 0.66 \cr
 DEC 5000-120  & R3000 &  20 & 64+64 &   1    &  50  & 16.4  & 0.91 & 0.73 \cr
 DEC 5000-25   & R3000 &  25 & 64+64 &   1    &  50  & 19.1  &      &      \cr
 Sony 3710     & R3000 &  20 & 64+64 &   1    &      & 12.6  & 1.00 & 0.80 \cr
 MIPS Magnum   & R3000 &  33 & 32+32 &   8    & 133  & 25.1  &      &      \cr
 HP/Apollo 705 & \PARISC &35 & 32+64 &   N/A  &      & 34.   &      &      \cr
 HP/Apollo 720 & \PARISC &50 & 128+256 & N/A  & 400  & 59.5  & 0.39 & 0.78 \cr
 HP/Apollo 730 & \PARISC &66 & 128+256 & N/A  & 528  & 76.8  &      &      \cr
 IBM 6000-320  & IBM   &  20 &  8+32 &   N/A  & 160  & 32.8  & 0.82 & 0.66 \cr
 IBM 6000-320H & IBM   &  25 &  8+32 &   N/A  & 200  & 41.2  &      &      \cr
 SUN 2         & Sparc &  40 &    64 &   N/A  &  49  & 24.7  & 0.72 & 1.15 \cr
 SUN ELC       & Sparc &  33 &    64 &   N/A  &  41  & 20.1  &      &      \cr
 Cray Y--MP    & Cray  & 166 &     0 &   N/A  &4200  &142.9  & 0.32 & 2.13 \cr
}}$$
\vskip 8pt
\noindent
{\bf Table 6.  Comparison of Current Workstations.} The Cray figures
are for a single CPU.  Only the MIPS R2000 and R3000 use a write through cache
architecture.  The fifth column does not apply to the other CPU types.
The last column shows how long it would take to reconstruct an E769
benchmark event if the computer clock speed were {\it scaled} to 25 MHz.
\eject
\leftline{\bf FIGURES}
\vskip 1in
\epsfysize=366pt
\centerline{\epsfbox{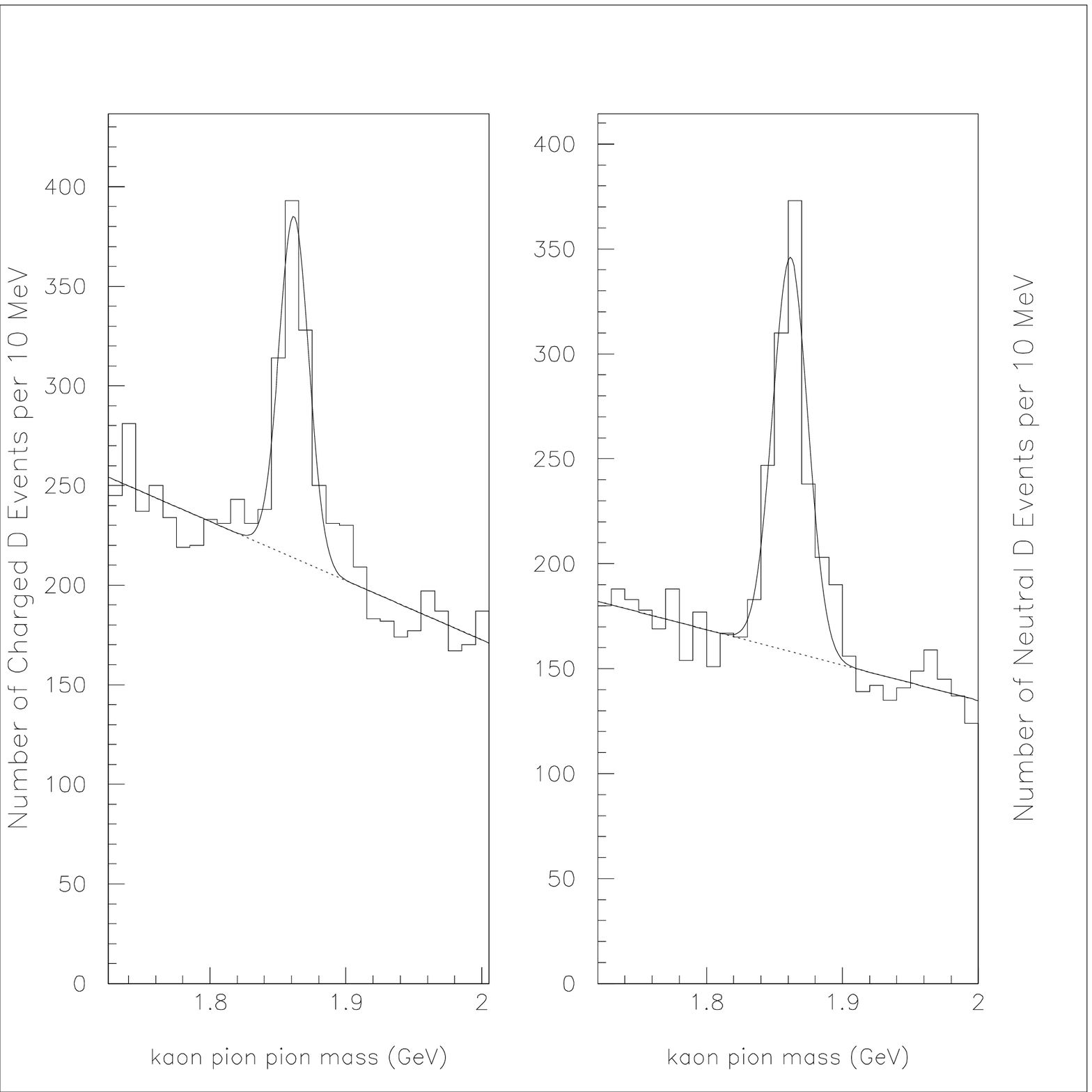}}
\leavevmode
\vskip .5in
\noindent
{\bf Figure 1.  Charm Meson Signals.}
$D^+(1.869) \rightarrow K^- \pi^+ \pi^+$  and
$D^0(1.865) \rightarrow K^- \pi^+$.

 
\epsfysize=366pt
\centerline{\epsfbox{sys.eps}}
\leavevmode
\vskip .5in
\noindent
{\bf Figure 2.  Network Configuration for Compute Servers and LAVC.} \quad
ETHERNET$^{TM}$ was used to connect four SGI 4D/240S compute servers and
a Local Area VAX cluster of VAXstations.  The VAXstations have 8-user
VAX/VMS licenses.
 
\vskip .5in
\epsfysize=244pt
\centerline{\epsfbox{bubble2.eps}}
\leavevmode
\vskip .5in
\noindent
{\bf Figure 3.  Structure of a CPS job.} \quad  Fermilab's Cooperative Process
Software [10] was used to distribute independent events from tape to CPUs
which processed the data.  CPS was then used to gather results together.
 
\end{document}